\documentclass[12pt]{article}
\usepackage{epsfig}
\usepackage{amsfonts}
\usepackage{latexsym}
\usepackage{amsmath,amssymb}
\usepackage{verbatim}
\usepackage{mathtools}
\usepackage{youngtab}
\usepackage{color}
\usepackage[font=small,labelfont=bf]{caption}
\usepackage{subcaption}

\numberwithin{equation}{section}

\newcommand{\bea}{\begin{eqnarray}}
\newcommand{\eea}{\end{eqnarray}}
\newcommand{\be}{\begin{equation}}
\newcommand{\ee}{\end{equation}}
\newcommand{\ba}{\begin{align}}
\newcommand{\ea}{\end{align}}
\newcommand{\bmat}{\begin{bmatrix}}
\newcommand{\emat}{\end{bmatrix}}

\renewcommand{\H}{\mathbb{H}}
\newcommand{\CC}{\mathbb{C}} 
\newcommand{\RR}{\mathbb{R}} 
\newcommand{\Z}{\mathbb{Z}} 

\newcommand{\scri}{{\cal I}}

  \makeatletter
  \let\over=\@@over \let\overwithdelims=\@@overwithdelims
  \let\atop=\@@atop \let\atopwithdelims=\@@atopwithdelims
  \let\above=\@@above \let\abovewithdelims=\@@abovewithdelims

\begin{document}

\begin{titlepage}
\begin{flushright}
NSF-KITP-12-165
\end{flushright}

\begin{center}
{\LARGE The Wave Function of Quantum de Sitter}

\vspace{6mm}
{Alejandra Castro${}^{a}$ and Alexander Maloney${}^{b}$ }\vspace{6.0mm}\\
\bigskip
{{\it $^a$Center for the Fundamental Laws of Nature, Harvard University, Cambridge 02138, MA USA}\vskip .5mm{\it $^b$Physics Department, McGill University, 3600 rue University, Montreal, QC H3A 2T8, Canada}
}
\medskip


\vspace{1cm}

\begin{abstract}
\noindent

We consider quantum general relativity in three dimensions with a positive cosmological constant.  The Hartle-Hawking wave function is computed as a function of metric data at asymptotic future infinity.  The analytic continuation from Euclidean Anti-de Sitter space provides a natural integration contour in the space of metrics, allowing us -- with certain assumptions -- to compute the wave function exactly, including both perturbative and non-perturbative effects.  The resulting wave function is a non-normalizable function of the conformal structure of future infinity which is infinitely peaked at geometries where ${\cal I}^+$ becomes infinitely inhomogeneous. We interpret this as a non-perturbative instability of de Sitter space in three dimensional Einstein gravity.

\end{abstract}

\end{center}

\setcounter{footnote}{0}

\end{titlepage}
\newpage

\baselineskip 17pt

\tableofcontents


\section{Introduction}

The application of quantum mechanics to cosmology is notoriously difficult.  Even basic notions -- such as unitary evolution on a Hilbert space -- are difficult to reconcile with cosmological features such as space-like singularities and inflationary expansion.  Perhaps the most notable proposal in this regard is that of Hartle and Hawking  \cite{Hartle:1983ai}, who argued that the universe is described by a standard quantum mechanical wave function which can be computed via a Euclidean path integral.  Without a complete understanding of quantum gravity, however, this proposed wave function is difficult to compute precisely.  The goal of the present paper is to investigate this wave function in 2+1 dimensions, where one can give a precise prescription for the computation of this path integral and understand its basic features.  

We will study 2+1 dimensional Einstein gravity with a positive cosmological constant: 
\be\label{action}
S =  {1\over 16 \pi G} \int_{M} d^3 x \sqrt{g} \left(R - {2\over \ell_{\rm dS}^2}\right)
\ee
The equations of motion state that space-time is locally, though not necessarily globally, three dimensional de Sitter space.   We wish to compute the wave function of the universe $\Psi(h)$, regarded as a functional of the topology and geometry (denoted collectively $h$) of a two dimensional spatial slice.  Although the theory contains no local degrees of freedom the wave-function $\Psi(h)$ still contains a great deal of interesting information about the global structure of space-time.  In particular, it is possible to find classical solutions (such as quotients of dS${}_3$) with arbitrary topology and metric $h$ on a given spatial slice; $\Psi(h)$ can be regarded as a wave function over this family of classical geometries.

Schematically, the evolution of the wave function $\Psi(h)$ should be determined by a Lorentzian path integral of the form
\be\label{psidef}
\Psi(h) = \int_{g|_{\partial M} = h} {\cal D} g~e^{i  S}~,
\ee  
where the integral is over a suitable space of three dimensional Lorentzian manifolds $M$.  The dependence on $h$ enters through the boundary values of the three metric $g$.  In addition, one must set some initial data which will fix the choice of state.
For example, in perturbative quantum field theory around a fixed de Sitter background one obtains the usual Bunch-Davies wave function by demanding that the metric fluctuations are purely positive frequency across a past event horizon. 
At the non-perturbative level, however, it is unclear precisely how one should fix the initial data in \eqref{psidef}.   The proposal of Hartle-Hawking \cite{Hartle:1983ai} is that the boundary conditions are determined by a particular analytic continuation to Euclidean signature.  More precisely, the wave function is given by a ``no boundary'' prescription: one integrates over Euclidean geometries whose only boundary is the spatial slice where the data $h$ are fixed.  According to the original proposal of  \cite{Hartle:1983ai}, each  such geometry is weighted by the action of general relativity in Euclidean signature.

The wave function $\Psi(h)$ described above is also a central object of interest in the conjectured dS/CFT correspondence \cite{Strominger:2001pn} (see also \cite{Maldacena:2002vr, Witten:2001kn}).\footnote{See \cite{Anninos:2012qw} for a more recent review and \cite{Alishahiha:2004md, McFadden:2009fg, Sekino:2009kv} for related approaches to holographic cosmology.} As the spacelike slice is taken towards $\scri^+$ the volume of the spatial slice 
diverges and the phase of the wave function oscillates rapidly.  In particular, near $\scri^+$ the wave function takes the form (see e.g. \cite{Maldacena:2002vr, Maldacena:2011mk})
\be\label{sct}
\Psi(h) \sim e^{i S_{ct}(h)} \Psi\left([h]\right)_{\rm ren}~.
\ee
Here $S_{ct}$ is a local functional of $h$ which diverges near $\scri^+$.  The  dependence of the wave function on non-local properties of $h$ is captured by the complex function $\Psi\left([h]\right)_{\rm ren}$.  In particular, $\Psi_{\rm ren}$ depends on the conformal class $[h]$ of the spatial metric;  all other information about the spatial metric is scaled away as one approaches $\scri^+$.   Equation (\ref{sct}) is the de Sitter analogue of the usual holographic renormalization procedure in Anti-de Sitter space (see \cite{Skenderis:2002wp} for a review).  
In the dS/CFT correspondence  $\Psi([h])_{\rm ren}$ is identified with the partition function of a Euclidean conformal field theory.   In the present case we have a purely metric theory of gravity, so consider only the dependence of $\Psi_{\rm ren}$ on the conformal structure $[h]$ of the geometry at $\scri^+$.  For theories of gravity containing additional bulk fields the partition function would depend on additional boundary data, and would be regarded as a generating function for correlation functions of the corresponding boundary operators.  

\begin{figure}
        \begin{subfigure}[c]{0.5\textwidth}
                \centering
                \includegraphics[width=\textwidth]{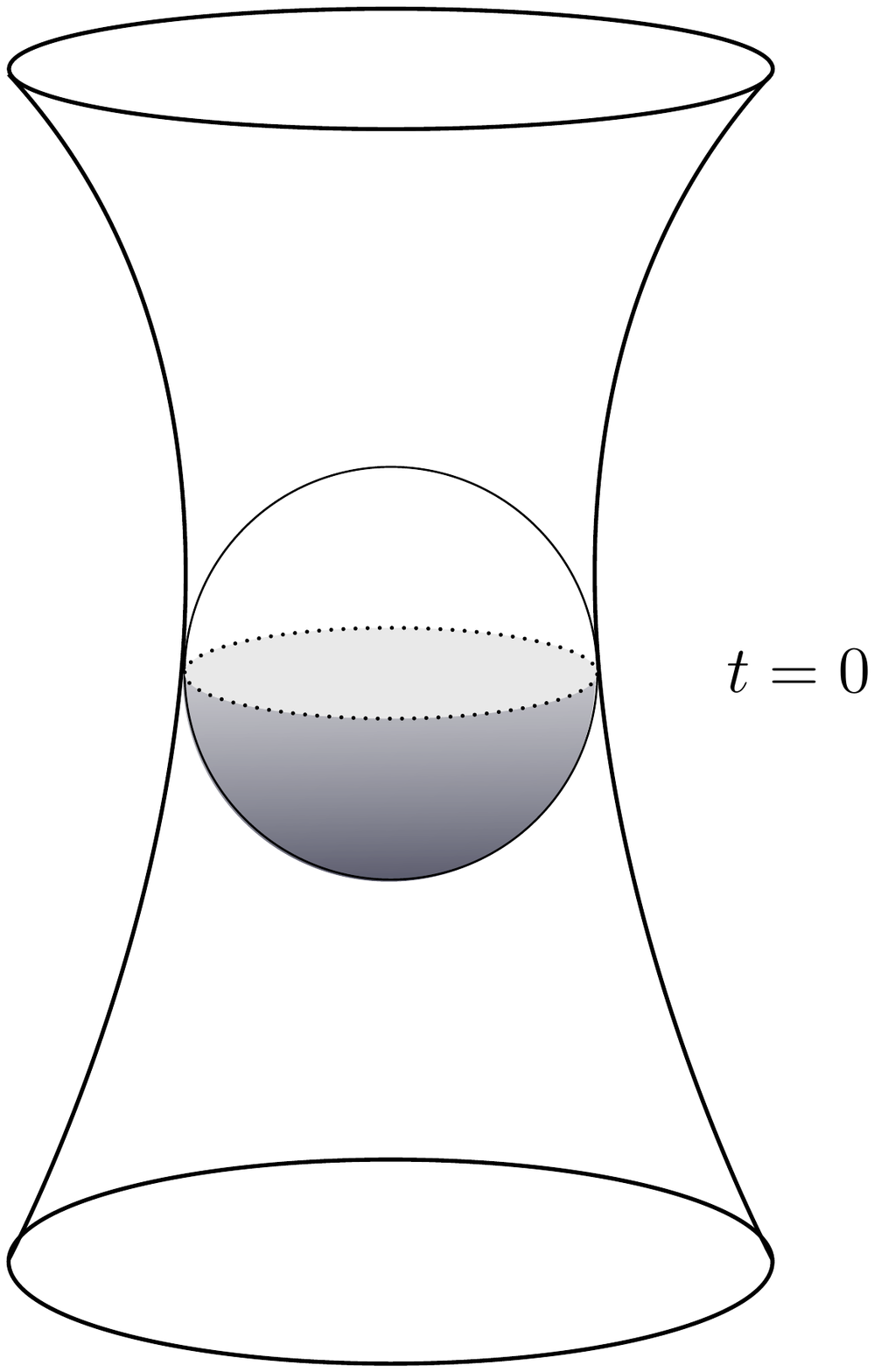}
                \caption{In the first type of analytic continuation, the matching between de Sitter space and the sphere happens on the equator at $t=0$. }
                \label{fig:a}
        \end{subfigure}%
          \quad
        \begin{subfigure}[c]{0.5\textwidth}
                \centering
                \includegraphics[width=\textwidth]{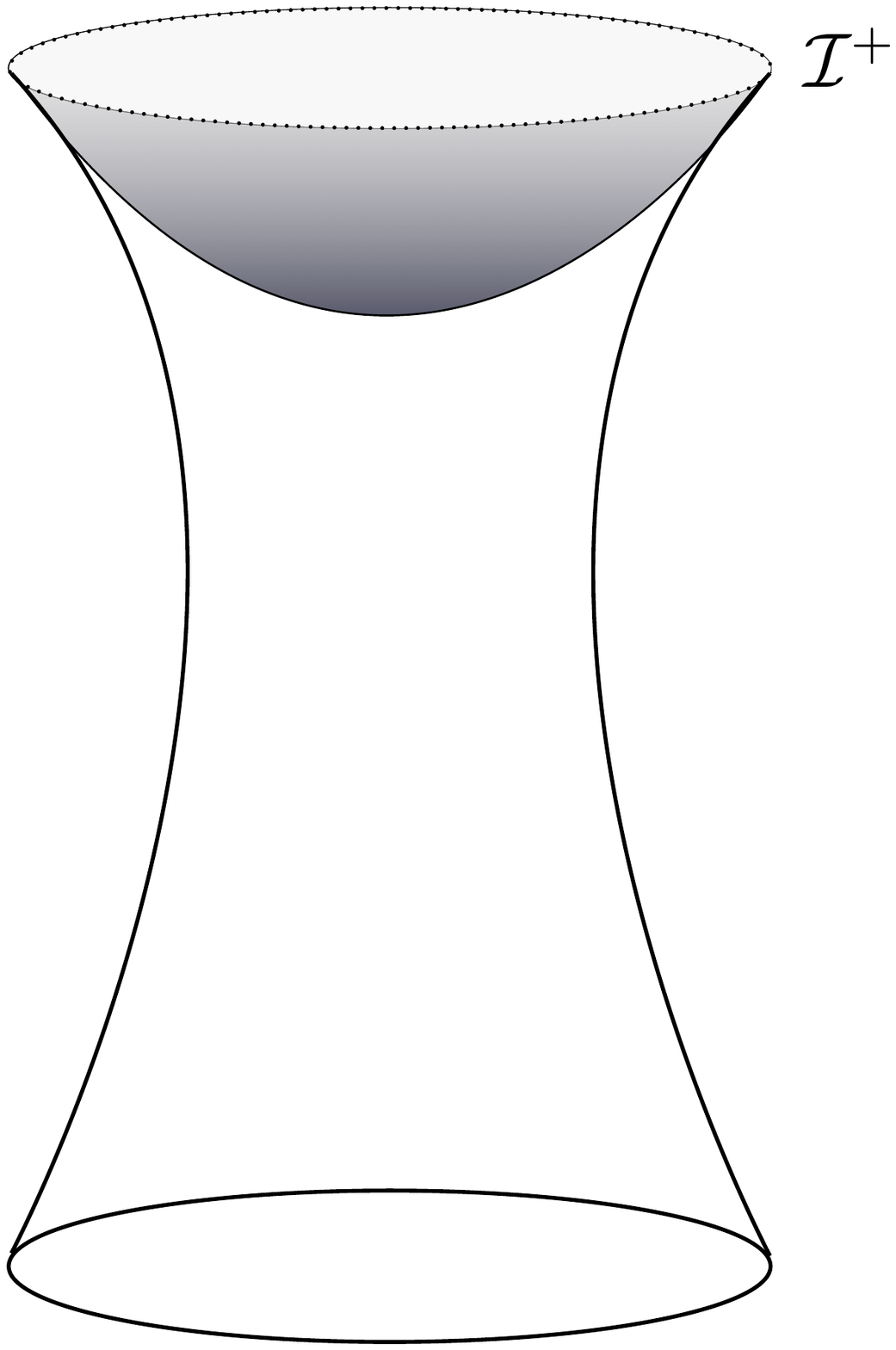}
                \caption{
                In the second type of analytic continuation, the matching between de Sitter space and Euclidean AdS (the hyperboloid) happens at $\scri^+$.}
                \label{fig:b}
        \end{subfigure}
        \caption{}\label{fig:c}
\end{figure}

Our goal is to compute $\Psi([h])$ directly in gravity, rather than via a conjectured boundary dual.\footnote{In the following discussion $\Psi([h])$ will refer to the renormalized wave function; we will omit the subscript in \eqref{sct}.}  We will use an analytic continuation to Euclidean gravity.  There are two distinct types of analytic continuation to Euclidean signature which could be used to do this.  The first analytic continuation procedure is that originally used by Hartle and Hawking  \cite{Hartle:1983ai} (see also \cite{Gibbons:1976ue}).  It relates the action \eqref{action} to that of general relativity in Euclidean signature with a positive cosmological constant.  At the level of perturbation theory this can be understood as a complex coordinate transformation which takes the global de Sitter solution into the sphere.  Explicitly, if $t$ is the global de Sitter time coordinate, then under the Wick rotation $t\to it$:
\be
ds^2 = \ell_{\rm dS}^2 (-dt^2 +\cosh^2 t d\Omega^2)  \to \ell_{\rm dS}^2 (dt^2 +\cos^2 t d\Omega^2)~.
\ee
The surface of vanishing extrinsic curvature at $t=0$  of the Lorentzian geometry matches onto the $t=0$ slice of the Euclidean geometry.  This is sketched in Figure (\ref{fig:a}).  In perturbative quantum field theory, the resulting Euclidean path integral leads to the usual Bunch-Davies vacuum state used to compute fluctuations in inflationary cosmology.  

For our purposes, however, the standard Hartle-Hawking approach has two drawbacks.  First, it is not known whether the full Euclidean gravity path integral with a positive cosmological constant can be defined and computed precisely.  Second, this procedure computes the properties of the state at finite time ($t=0$ in Figure (\ref{fig:a})).   In order to compute $\Psi([h])$ one would need to evolve this wave function to the future boundary at $t\to\infty$ using a Lorentzian path integral of the form \eqref{psidef}, which we likewise do not know how to compute precisely.  It would be preferable to use an analytic continuation procedure which circumvents this intermediate step and allows us to compute $\Psi([h])$ directly via a Euclidean computation.  Indeed, the holographic principle suggests that perhaps asymptotic quantities such as $\Psi([h])$ can be well defined even if $\Psi(h)$ itself is not well defined at finite time.

Both of these difficulties are resolved by the elegant analytic continuation procedure proposed by Maldacena \cite{Maldacena:2002vr}.\footnote{See also \cite{McFadden:2009fg, Harlow:2011ke,Maldacena:2011mk} and references therein.} The starting observation is that the analytic continuation $z \to -iz$, where $z$ is the planar time coordinate, along with $\ell_{\rm AdS}\to i \ell_{\rm dS}$ takes de Sitter space into Euclidean Anti-de Sitter space (i.e. the hyperboloid):
\be\label{mhh}
ds^2 = \ell^2_{\rm AdS} \left({-dz^2 + d\vec x^2 \over z^2}\right) \to \ell^2_{\rm dS} \left({dz^2 + d\vec x^2 \over z^2}\right)~.
\ee
Moreover, the usual Bunch-Davies boundary condition at the past horizon $z\to-\infty$ becomes, upon analytic continuation, the condition that modes are normalizable in the interior of Euclidean AdS.  Thus the wave function $\Psi(h)$ is computed in terms of a path integral for Euclidean gravity with a {\it negative} cosmological constant.  
An important feature of this continuation is that the future boundary of the Lorentzian de Sitter geometry coincides with the asymptotic boundary of Euclidean AdS; they are both given by $z=0$ in (\ref{mhh}).  This is indicated in Figure (\ref{fig:b}).  This analytic continuation procedure computes the wave function at $\scri^+$ directly, circumventing the need for any Lorentzian computation.   Under this analytic continuation, equation (\ref{sct}) becomes precisely the usual relation between the partition function of bulk gravity in Anti-de Sitter space and that of a CFT, including the usual local counterterms. Thus this method is well adapted to the computation of the boundary wave function $\Psi([h])$.  

The  analytic continuations described above can be regarded as two attempts to define a path integral via a choice of contour of integration through the space of complex metrics.  Without a more detailed understanding of this path integral\footnote{Such as that described in \cite{Witten:2010cx, Harlow:2011ny} for certain relatively simple theories.} it is not clear which of these prescriptions is correct.   At the level of perturbation theory around a fixed de Sitter background, however, both analytic continuation prescriptions appear to define the same state, at least for general relativity without additional degrees of freedom \cite{Maldacena:2002vr}.  The  question is whether they differ at the non-perturbative level.  Indeed, it is not clear that the original Hartle-Hawking path integral can even be defined precisely beyond perturbation theory around a fixed classical background.\footnote{See \cite{Castro:2011xb, Castro:2011ke} for an attempt to do so in three dimensional gravity.}  Path integrals in Euclidean AdS are better understood.  For three dimensional gravity with certain boundary conditions they can be computed exactly.  Given that the Euclidean AdS results match those of the usual Hartle-Hawking contour at the perturbative level -- and can be defined precisely at the non-perturbative level -- we will therefore use the Euclidean AdS contour in the computation that follows.

 Our primary technical tool is the computation of the partition function of Euclidean AdS gravity by \cite{Maloney:2007ud}, which followed earlier work \cite{Brown:1986nw, Maldacena:1998bw, Dijkgraaf:2000fq, Witten:2007kt}.  The goal of the present paper is to reinterpret these results in the context of quantum cosmology.\footnote{In \cite{Maloney:2007ud} the authors argued that the Euclidean AdS gravity path integral could not be interpreted as the computation of a canonical ensemble partition function in Lorentzian AdS.  This conclusion was based on the traditional Hartle-Hawking analytic continuation prescription which relates Euclidean AdS to Lorentzian AdS.  In the present paper we reinterpret the analysis of \cite{Maloney:2007ud} as defining a state in Lorentzian de Sitter space, using Maldacena's analytic continuation prescription. Thus the conclusions are quite different.}


It is worth emphasizing that for certain theories of gravity the analytic continuation from de Sitter space to the hyperboloid is much more natural than that from de Sitter space to the sphere.  The isometry group of de Sitter space, $SO(3,1)$, is identical to that of Euclidean AdS.   If gravity is formulated as a gauge theory -- as, for example, in the Chern-Simons formulation of 3D gravity -- $SO(3,1)$ appears as the group of local gauge symmetries \cite{Achucarro:1987vz,Witten:1988hc}.  In a gauge theory, the usual Euclidean continuation prescription involves a complex change of coordinates but not a change of the gauge group.  This indicates that the analytic continuation from dS to Euclidean AdS is in a sense the correct one, if  one formulates gravity as a local gauge theory.\footnote{We note, however, that the Chern-Simons formulation of 3D gravity does not appear to capture all of the desired features of quantum gravity, in particular black hole entropy \cite{Witten:2007kt}. Thus it is not clear how seriously one should take this argument.} The sphere, on the other hand, has isometry group $SO(4)$ and is only obtained if one in addition changes the gauge group.    

In addition to three dimensional Einstein gravity, there is an second theory to which this observation applies:  Vasiliev's higher spin theory of gravity in four dimensions \cite{Vasiliev:1990en}.    In particular, Vasiliev theory with a positive cosmological constant has dS${}_4$ as its maximally symmetric Lorentzian solution, and Euclidean AdS${}_4$ as its maximally symmetric Euclidean solution.  Much like the Chern-Simons description of three dimensional gravity, the theory is formulated purely in terms of connection variables.  Indeed, recently there has been a great deal of progress on the formulation of the   dS/CFT correspondence for higher spin theories of gravity \cite{Anninos:2011ui, Ouyang:2011fs,Ng:2012xp,Das:2012dt}, based in part on the analogy with the corresponding theories in AdS.  The wave function of Vasiliev gravity with future de Sitter boundary was studied using the conjectured boundary gauge theory dual \cite{Anninos:2012ft}, and shown to exhibit similar features to those we will discover below.  

This discussion highlights, however, an important drawback of the present approach, which is that it only applies to certain special theories of gravity. For generic matter coupled to gravity, the analytic continuation (\ref{mhh}) makes the solution complex. Without a more detailed understanding of the path integral, it is not clear that such complex solutions should be regarded as genuine saddle point contributions to an appropriate path integral.\footnote{See \cite{Hertog:2011ky, Hartle:2012qb} for a discussion of the contribution of such saddles.}  The Maldacena contour should therefore be applied with caution.


In the next sections we describe in more detail the exact computation under consideration, where the wave function is computed explicitly for geometries which asymptote to a torus at future infinity.  We first describe the family of Lorentzian geometries on which this wave function is supported -- essentially de Sitter analogues of the BTZ black hole -- and then describe the explicit computation of the wave function.  We will conclude that the wave function is non-normalizable, which is interpreted as a quantum mechanical instability of the de Sitter vacuum.

\section{The Classical Geometries}\label{sec:geom}

The locally de Sitter solutions of Einstein gravity with a positive cosmological constant can have arbitrary topology at future infinity.
  In the maximally symmetric solution -- global de Sitter space  -- $\scri^+$ has the topology of the sphere $S^2$.  For quotients of de Sitter space, however, the topology can be more interesting.  The wave function $\Psi([h])$ depends on the conformal structure of the metric $h$ at future infinity; for the sphere this conformal structure is completely fixed.  Thus on global de Sitter space there is not much to compute, at least in a theory of pure gravity without additional fields; the wave function on the sphere is just a number.  To obtain interesting dynamical information about the wave function of the universe we should consider more complicated topologies for which the moduli space of conformal structures is non-trivial. 

In the semi-classical limit, the wave function $\Psi([h])$ should be regarded as a wave function over geometries which solve the equations of motion.  We will first describe these classical geometries.  We will consider in detail the case where future infinity has the topology of a torus $T^2$, before commenting at the end of this section on the more general case.
The conformal data of $T^2$ is encoded in a complex parameter $\tau$, so we denote the corresponding wave function $\Psi(\tau, \bar \tau)$.\footnote{We use this notation to emphasize that the wave function of Einstein gravity is neither a holomorphic function of $\tau$ nor, necessarily, the square of one.  In order to obtain a holomorphic function of $\tau$ one would have to consider some version of topologically massive gravity or chiral gravity in de Sitter space, as discussed in e.g. \cite{Castro:2011ke}.}  

The classical geometries for which future infinity has $T^2$ topology are quotients $dS_3/\Z$.
To see this simply, let us write the metric on  de Sitter space in cylindrical coordinates\footnote{These coordinates are related by analytic continuation $r\to i r$, $\theta \to i \theta$ to the usual coordinates on the static patch.} 
\be\label{tmetric}
ds^2 = -dt^2 + \cosh^2t d\theta^2 + \sinh^2 t d\phi^2~.
\ee
In global de Sitter space $\phi\sim\phi+2\pi$ and $\theta$ runs from $-\infty$ to $\infty$.  We can further identify $\theta\sim\theta+2\pi$ to obtain a geometry which approaches a torus at future infinity $t\to\infty$; this is an example of a quotient dS${}_3/\Z$.  The surface $t=0$ is a coordinate singularity in global de Sitter space, but in the quotient becomes a genuine cosmological singularity of Milne type. The wave function $\Psi(\tau, \bar \tau)$ is supported on a family of geometries with big-bang singularities of Milne type.

In fact there are an infinite number of such saddle point geometries dS${}_3/\Z$ which generalize \eqref{tmetric}, which we will now construct.\footnote{These geometries have been considered before in the context of dS/CFT (see e.g. \cite{Balasubramanian:2001nb}).} They will be labelled by elements of the coset $\Z\backslash SL(2,\Z)$.  
We  start from the observation that Lorentzian dS$_3$ is the quotient $SL(2,\CC)/SL(2,\RR)$. An element of this quotient can be decomposed as
\be
g=h (h^*)^{-1}~,\quad h\in SL(2,\CC)~.
\ee
From here it is clear that $g$ is invariant under $h\to h \alpha$ with $\alpha \in SL(2,\RR)$. The action of $SL(2,\CC)$  on $g$ and $h$
\be\label{sl2}
h\to \lambda h ~, \quad g\to \lambda g (\lambda^*)^{-1} ~, \quad \lambda\in SL(2,\CC)~,
\ee
leaves the metric
\be
ds^2=-{\ell^2_{\rm dS}\over2}{\rm Tr}(g^{-1}dg\,g^{-1}dg)~
\ee
invariant.

The inflationary (i.e. planar) patch of de Sitter space is obtained by choosing  coordinates for $g$
\be\label{poincare}
g=\left[
\begin{array}{cc}
w/z  & z- w\bar w/z\\
-1/z&\bar w/z
\end{array}\right]~,
\ee
with $z\in\RR$ and $w\in \CC$. The metric is
\bea\label{dspoinc}
ds^2=
 {\ell^2_{\rm dS}\over z^2}\left(-dz^2+dwd\bar w\right)~.
\eea

In this language it is easy to construct the quotients dS$_3/\Z$.  We make the identification
\be\label{quot}
g\sim \gamma g (\gamma^*)^{-1}~,
\ee
with
\be
\gamma=\left[
\begin{array}{cc}
e^{-i\pi\tau} &0\\
0&e^{i\pi\tau}
\end{array}\right]~,\quad \tau=\tau_1+i\tau_2~,
\ee
where $(\tau_1,\tau_2)$ are real.\footnote{The element $\gamma$ here is exactly the same as the one that defines Euclidean BTZ. See appendix \ref{app:BTZ} for further details on the relation between this construction and that of the BTZ black hole. } In the coordinates \eqref{poincare}, the identification \eqref{quot} implies
\be\label{idents}
z\sim e^{-i\pi (\tau-\bar\tau)} z~,\quad w\sim e^{-2\pi i\tau}w~.
\ee
A more convenient set of coordinates is given by
\bea\label{change1}
z&=&\left(|\tau|^2\over t^2-\tau_1^2\right)^{1/2}\exp(\tau_2\phi+\tau_1 \theta)~,\cr
w&=&\left(t^2+\tau_2^2\over t^2-\tau_1^2\right)^{1/2}\exp(-i\tau(\phi+i \theta))~,
\eea
where now the action of \eqref{quot} on the coordinates $(t,\phi,\theta)$ is
\be\label{aa:1}
\hat w=\phi+i \theta \sim \hat w +2\pi \sim \hat w + 2\pi \tau~.
\ee
In these new coordinates the metric is
\be\label{aa:2}
{ds^2\over \ell^2_{\rm dS}}=-{t^2dt^2\over(t^2-\tau_1^2)(t^2+\tau_2^2) }+t^2\left(d\phi -{\tau_1\tau_2\over t^2}d\theta\right)^2+{(t^2+\tau_2^2)(t^2-\tau_1^2)\over t^2}d\theta^2~.
\ee
This describes a big bang/crunch geometry, with Milne singularity at $t=\tau_1$. The conformal class of the spacelike boundary at $\scri^+$ ($t \to \infty$) is a torus with conformal structure parameter $\tau$. 
We note that this geometry has no static patch.  
 
Equation \eqref{aa:2} describes one possible geometry on which the wave function $\Psi(\tau, \bar \tau)$ is supported. In fact there are an infinite number of such classical geometries, each of which has a boundary torus with the same conformal structure $\tau$ at future infinity.  These geometries are related to \eqref{aa:2} by modular transformation.  To see this, note that the $\phi$ and $\theta$ coordinates are not treated democratically in the geometry (\ref{aa:2}). In this construction we made a choice of contractible cycle; in the coordinates \eqref{aa:2}   the $\theta$ circle shrinks to zero size at $t=\tau_1$, but $\phi$ does not.  The other geometries will correspond to other cycles which are made contractible at the Milne singularity.
These geometries will be related to \eqref{aa:2} by the action of the modular group $SL(2,\Z)$, which is the mapping class group of the torus.

We can construct these geometries explicitly by performing a large diffeomorphism that acts non-trivially on the boundary.  The modular transformation
%
%
\be\label{g:ba}
\tau \to \gamma\tau ={a \tau + b\over c\tau + d}~, \qquad \gamma = \left({a~~b\atop c~~d}\right) \in SL(2,\Z)~,
\ee
 is implemented by the diffeomorphism
 \be\label{coord}
 w~\to~ {w\over c\tau+d}~.
 \ee
 This provides a family of geometries labelled by $SL(2,\Z)$.  We note that shifting $\tau \to \tau+1$ does not change the identifications \eqref{idents}.  Thus distinct geometries should be labelled not by $SL(2,\Z)$ but by elements of the coset $\Z \backslash SL(2,\Z)$, where $\Z$ is the subgroup generated by $T = \left({1~1\atop0 ~ 1} \right): \tau \to \tau+1$.
 
 We emphasize that, although these new solutions are related by the coordinate transformation \eqref{coord}, they are regarded as distinct contributions to the path integral. We adopt the definition that diffeomorphisms whose action on the boundary is non-trivial are physical, and that only those diffeomorphism that act trivially at future infinity are truly pure gauge.  This definition was first discussed in the original proposal of the dS/CFT correspondence \cite{Strominger:2001pn}.  Further, following the construction of conserved charges in  the dS/CFT correspondence (see e.g. \cite{Klemm:2001ea, Bousso:2001mw}), it is straightforward to check that these modular transformation act non-trivially.     
%

An important observation is that these geometries are an incomplete classification of geometries whose boundary is a two-torus. Since we are not restricting the classical phase space to include only smooth Lorentzian solutions, there is a priori no reason to exclude more general quotients. For example,  quotients of dS$_3$  by $\Z\times \Z_m$ (with $m$ an integer) satisfy our boundary conditions, and the additional action of $\Z_m$ creates a conical singularity on top of the Milne singularity.  However, as we will see below it is the quotients by $\Z$ which continue naturally to the Euclidean BTZ black hole solutions via Maldacena's analytic continuation.  Thus it is most natural to regard the wave function as supported on the quotients $dS_3/\Z$.  We will elaborate  on this point in the next section. 

Given the above discussion, it is clear at least in principle how one can construct quotients of $dS_3$ for which $\scri^+$ has more complicated topology.  In global de Sitter space the $SL(2,\CC)$ isometries \eqref{sl2} act naturally on the boundary as conformal transformations.  These are the usual Mobius transformations of the $S^2$ at future infinity.  To obtain a non-trivial Riemann surface at $\scri^+$ one quotients de Sitter space by a discrete subgroup of $SL(2,\CC)$, i.e. a Kleinian group.  By uniformization, any compact Riemann surface with arbitrary conformal structure can be obtained by such a quotient of the Riemann sphere $S^2$.   Equation \eqref{sl2} tells us how to extend this identification into the bulk to obtain a quotient of $dS_3$.   The conformal moduli of the Riemann surface then enter as the continuous parameters that label the choice of discrete subgroup. The wave function is then regarded as a function on these geometries. 

As in the torus case, there will be many locally de Sitter geometries whose boundary is given by the same Riemann surface.  Indeed, one can act on each such geometry with the mapping class group (the higher genus analogue of the modular group $SL(2,\Z)$) to obtain an infinite family of geometries each of which asymptote to the same conformal geometry at future infinity.   While it is difficult to write the metrics on these geometries explicitly, they will contain at the very least Milne singularities where cycles in the boundary geometry shrink to zero size.

\section{The Wave Function}

We now proceed to compute the wave function of the universe using Maldacena's analytic continuation prescription.  This will allow us to compute both perturbative and non-perturbative corrections to the usual Euclidean vacuum state. We will focus on the case where the boundary at future infinity is a two torus. 
As the conformal structure of the torus is modular (i.e. $SL(2,\Z)$) invariant, $\Psi(\tau, \bar \tau)$ will be likewise be an $SL(2,\Z)$ invariant function of $\tau$.  It can be viewed as a function on the fundamental domain of the action of $SL(2,\Z)$ on the upper half $\tau$-plane.


According to Maldacena's analytic continuation prescription  \cite{Maldacena:2002vr}, a correlation function on $\scri^+$ of massless fields will equal precisely the answer obtained by analytic continuation from Euclidean AdS where we take
the central charge 
\be
c_{\rm AdS}={3 \ell_{\rm AdS} \over 2 G} \to ic_{\rm dS}=i {3 \ell_{\rm dS} \over 2 G}
\ee
to be pure imaginary. 

Under this analytic continuation Lorentzian gravity with a positive cosmological constant becomes Euclidean gravity with a negative cosmological constant.  The wave function $\Psi(\tau, \bar \tau)$ becomes the partition function of three dimensional Euclidean AdS gravity with torus boundary conditions, which we denote $Z(\tau, \bar \tau)$.  This partition function was computed in explicitly in \cite{Maloney:2007ud}
following earlier work \cite{Brown:1986nw, Maldacena:1998bw, Dijkgraaf:2000fq, Witten:2007kt}. The result is%
\be\label{cc:3}
{Z(\tau,\bar\tau) = \sum_{(c,d)=1\atop c\geq0} Z_{0,1}(\gamma
\tau,\gamma\bar\tau),}
\ee
 where 
 \be
 {\gamma \tau = {a\tau+b\over
c\tau+d}~,~~~~~~~~~\gamma = \left(a\ b \atop c\ d\right) \in
SL(2,\Z)}~,
\ee
 and 
\be\label{z01}
{Z_{0,1}(\tau,\bar\tau) = \left|q^{-k}
\prod_{n=2}^{\infty}(1-q^{n})^{-1}\right|^{2} = {|\bar q
q|^{-k+1/24}|1-q|^{2}\over |\eta(\tau)|^{2}}}~.
\ee
Here $\eta$ is the Dedekind eta function. The coupling $k=c_{\rm AdS}/24$ is real for AdS gravity. 

Let us now describe the various terms in this expression and their interpretation in dS gravity.
The summation in equation \eqref{cc:3} is over all smooth Euclidean solutions which have a torus at their asymptotic boundary.   These are the $SL(2,\Z)$ generalizations of the BTZ black hole, whose geometries are reviewed in appendix \ref{app:BTZ}.   It is this sum which guarantees that the partition function $Z(\tau, \bar \tau)$ is a modular invariant  function of $\tau$. 

The sum over $(c,d)$ in \eqref{cc:3} can be regarded a sum over the coset $\Z\backslash SL(2,\Z)$, where $\Z$ is the subgroup generated by $T=\left({1~1\atop 0~1}\right)\in SL(2,\Z)$.  Geometrically, $T$ is a Dehn twist which leaves the bulk geometry unchanged.  In the BTZ language, $T$ is a modular transformation which does not change the mass and angular momentum of the black hole.  In any case, the summand $Z_{0,1}$ is invariant under $T$, so only by considering the sum over the coset $\Z\backslash SL(2,\Z)$ can one obtain a finite sum.  
We emphasize that this coset labels $all$ smooth solutions to the equations of motion with a torus at the boundary. Thus all smooth saddle points which give non-perturbative contributions to the wave function are included.\footnote{In principle, it may be possible to alter the definition of our path integral to include non-smooth Euclidean saddle points.  Without a precise definition of the path integral we have no way to know for sure.  However, we do not expect that our qualitative result -- the divergence of the wave function at boundaries of moduli space -- will be altered.}  

$Z_{0,1}$ is the partition function of the thermal AdS saddle, including all perturbative (loop) corrections.  The $|q|^{-2k}$ term in \eqref{z01} is the regularized classical action of Anti-de Sitter space, and the remaining terms describe a one loop correction to this action.
The boundary conditions which define the de Sitter wave function in the far past are precisely those obtained by analytic continuation from the requirement of regularity in the interior of Euclidean AdS.  Hence, around the vacuum solution the perturbative corrections to the wave function map precisely to the perturbative corrections to the classical partition function given by equation \eqref{z01}.  In Euclidean AdS gravity these perturbative corrections can be computed in two different ways.  The method of reference  \cite{Maloney:2007ud} involved a direct quantization of the appropriate phase space of metrics.  The corrections can also be computed using a more traditional one-loop determinant \cite{Giombi:2008vd}.  In either case, the answer is  \eqref{z01}.  

There is in principle a third method, which is to simply compute the corrections to the Hartle-Hawking state for perturbative gravitons directly in de Sitter space, as was done in \cite{Maldacena:2002vr}; this would presumably resemble closely the computation of \cite{Giombi:2008vd}.  We will focus on the Euclidean AdS construction, as only in this case are we able to include non-perturbative corrections.

Indeed, the analytic continuation $c_{\rm AdS}\to ic_{\rm dS}$ allows us to identify the wave function $\Psi(\tau, \bar \tau)$ as the analytic continuation of the partition function $Z(\tau, \bar \tau)$ at the non-perturbative level.  
This is not merely a convenient computational tool. It provides a precise definition of the wave function via the Maldacena contour. In fact, the sum over geometries \eqref{cc:3} can now be interpreted in Lorentzian signature as a sum over dS$_3/\Z$ cosmologies given by \eqref{aa:2}.  Each of the Lorentzian geometries described in the previous section will analytically continue to a smooth Einstein manifold in Euclidean AdS.  Thus each of the terms in the sum \eqref{cc:3} can be interpreted directly as a quantum mechanical amplitude associated to a particular cosmology.

In principle one should worry that the form of these non-perturbative corrections is subtle, in that the answer  differs depending on whether we analytically continue in $c_{\rm AdS}$ before or after performing the sum over geometries.  From the explicit results below, however, this does not seem to be the case.  We will therefore perform the sum first and then analytically continue, using the results of \cite{Maloney:2007ud}.   

We first consider the analytic continuation of the leading perturbative piece  \eqref{z01}  when $c_{\rm AdS}\to i c_{dS}$.  This means that $k=i c_{dS}/24$ is purely imaginary.  In  the usual perturbative regime where $\tau = \tau_1+i\tau_2 \to i \infty$ the partition function $Z_{0,1}$ diverges like 
\be\label{pertdiv}
Z_{0,1}\sim |q|^{-2k}(1+\cdots) = e^{4 \pi k \tau_2 }(1+\cdots)~,
\ee
when $k$ is real.  This becomes a pure phase when $k$ is imaginary.  Thus the classical contribution of $Z_{0,1}$ to the wave function $\Psi(\tau, \bar \tau)$ is a pure phase at $\tau \to i \infty$.  However, the next to leading piece also diverges at $\tau_2 \to 0$, because at this point 
\be\label{newdiv}
|Z_{0,1}(i\tau_2)|\sim  {|1-e^{-2\pi \tau_2}|^2 \over |\eta (i \tau_2)|^{2}} \sim  e^{\pi/(6 \tau_2 )}  ~.
\ee
Thus even though the perturbative piece is finite at the ``usual" pole at $\tau=i\infty$, it has a divergence at the modular image of this pole $\tau=0$.  In Euclidean AdS (with real central charge) this divergence is of no concern; it is much smaller than a stronger divergence which will occur due to one of the other saddles (the Euclidean BTZ saddle) at $\tau=0$.  In the Euclidean AdS picture this divergence can be interpreted as a one loop renormalization of the vacuum energy of the ground state.  In de Sitter space, however, this divergence is  crucial.  Unlike the classical divergence \eqref{pertdiv},  there are no factors of $k$ in \eqref{newdiv}.  Thus while the usual tree level divergence \eqref{pertdiv} becomes a phase under analytic continuation the one loop piece does not.  

In fact, because the Dedekind eta function is (almost) modular invariant we can argue that $Z_{0,1}$ will have a pole at every rational point on the real $\tau$ axis.  This is because in \eqref{z01} the $|q|^{1/12}$ factor cancels the pole in the eta function at $\tau=i\infty$, but not at any of the modular images of $i\infty$.  Based on this, it is clear what should happen when we sum over geometries; we will find that the wave function has a pole at the cusp $\tau=i\infty$, because of analogous one-loop terms which appear in the expansion around other (modular transformed) geometries.  

In order to confirm this expectation, we must discuss an additional subtlety. The partition function $Z(\tau,\bar \tau)$ is formally divergent due to the sum over  $\Z \backslash SL(2,\Z)$.  However, this sum can be regulated in a natural modular invariant way.  The result is complicated but can be evaluated explicitly in some cases.  To write the answer in a simple way, let us note that $Z(\tau, \bar \tau)$ is periodic under $\tau\to \tau+1$.  So we can Fourier expand in $\tau_1$.  Let us just consider the $\tau_1$ independent piece.  Then completely explicit expressions can be found for the wave function following the derivations in \cite{Maloney:2007ud}; see appendix  B for explicit formulas.  In particular if we expand near $\tau=i \infty$, the terms independent of $\tau_1$ are given by
\be
\Psi = Z_{0,1} + {1\over |\eta|^2} \sum_{n=0}^\infty c_n(ik) \tau_2^{-n}~,
\ee
where $Z_{0,1}$ is the analytic continuation of \eqref{z01} to imaginary $k$, and  $c_n$ are constants which can be computed explicitly.  They are polynomials in $k$ of order $n$.  For example
\be
c_0 = -6 , ~~~~~ c_1 = {(\pi^{3}-6\pi)(11+24k)\over 9 \zeta(3)} ~. 
\ee
As anticipated above, the wave function has a divergence at $\tau \to i \infty$ of the form 
\be
\Psi(\tau, \bar \tau) \sim -6\, \exp\left\{{\pi \over 6}\, {\rm Im}\, \tau \right\} ~~~~~{\rm as}~~~ {\rm Im}\, \tau \to \infty~.
\ee 
This divergence is precisely the one anticipated above; the factor of $-6$ is due to the regularization of the sum over modular images.    We emphasize that $Z_{0,1}$, which includes the classical action of the usual (thermal AdS) saddle along with its perturbative corrections, is finite. The divergence comes from non-trivial saddles, and in fact from  the one-loop contribution to the action of these saddles. 

We conclude that the wave-function of the universe has a peak at very ``asymmetric" universes where ${\rm Im}\, \tau$ is large.\footnote{Recall that ${\rm Im}\, \tau$ is essentially a measure of the inhomogeneity of the torus.  A rectangular torus with cycles of length $1$ and $\tau_2$ has $\tau = i \tau_2$.} This peak renders the wave function non-normalizable.\footnote{This agrees with the computation of \cite{Castro:2011xb}, in which the wave function of dS${}_3$ was considered using a more standard Hartle-Hawking path integral over Euclidean geometries with positive curvature.  The advantage of the present computation is that we now have a physical interpretation of this divergence.} Thus the probability density is infinitely peaked at infinitely inhomogeneous universes.
We note that this peak is {\it exponential} in ${\rm Im}\, \tau$ as ${\rm Im}\, \tau\to\infty$, thus the wave-function is non-normalizable even with standard $SL(2,\RR)$ invariant measure on the upper half $\tau$ plane.     It is also interesting that the strength of the peak (the coefficient in the exponential) is independent of the de Sitter radius.  This is because it is essentially a quantum mechanical effect -- due to a one loop effect, albeit around a ``non-perturbative" saddle -- rather than a classical one.

Our interpretation is that three dimensional de Sitter gravity is quantum mechanically unstable.  The wave function  resembles that of a quantum mechanical system with a potential which is unbounded below.  It is tempting to speculate that this mechanism is a simple version of the more complicated instabilities of de Sitter vacua observed in string theory constructions (see e.g. \cite{Kachru:2003aw}).\footnote{Our instability might even be related to the instabilities of de Sitter discussed in \cite{Banks:1984np, Polyakov:2009nq, Polyakov:2012uc}.}
In order to make this analogy precise, however, one would want to determine the endpoint of the instability.  In the present case the answer to this is unclear.   It might be necessary to add local degrees of freedom in order for the theory to have a sensible true vacuum ground state.  Vasiliev's theory of higher spin gravity is a  concrete setup to explore these issues \cite{Anninos:2012ft}. 

Of course, another natural direction is to explore to what extent these results can be generalized within dS$_3$ gravity. Consider, for example, the wave function with more complicated asymptotics.  When the boundary has higher genus, the corresponding Euclidean geometries are quotients of Euclidean AdS by Kleinian groups, and the resulting partition functions are more difficult to write down (see e.g. \cite{Krasnov:2000zq, Yin:2007gv}).  It seems likely that in this case that the wave function will be infinitely peaked at the boundaries of moduli space where the geometry of $\scri^+$ is infinitely inhomogeneous.  It would be interesting to investigate this further.


\section*{Acknowledgements}

We are grateful to Dio Anninos, Frederik Denef, Daniel Harlow, James Hartle, Tom Hartman, Thomas Hertog, Nima Lashkari, Don Marolf and Edward Witten and especially to Juan Maldacena, Stephen Shenker and Andy Strominger for useful conversations.  This work is supported by the National Science and Engineering Research Council of Canada, and the NSF under Grant No. PHY11-25915. AC's work is also supported by the Fundamental Laws Initiative of the Center for the Fundamental Laws of Nature, Harvard University. 

\appendix


\section{Euclidean BTZ}\label{app:BTZ}

In this appendix we review the construction of the Euclidean BTZ black hole as the quotient $ \H_3/\Z $. This should be compared with the analogous construction in section \ref{sec:geom} of $dS_3/\Z$. Euclidean AdS${}_3$ (i.e. $\H_3$) is the quotient $SL(2,\CC)/SU(2)$. An element of this quotient can be decomposed as
\be
g=h h^\dagger~,\quad h\in SL(2,\CC)~.
\ee
From here it is clear that $g$ is invariant under $h\to h \alpha$ with $\alpha \in SU(2)$. $SL(2,\CC)$  acts on $g$ and $h$ as
\be
h\to \lambda h ~, \quad g\to \lambda g \lambda^\dagger ~, \quad \lambda\in SL(2,\CC)~.
\ee
The metric is
\be
ds^2={\ell^2\over2}{\rm Tr}(g^{-1}dg\,g^{-1}dg)~.
\ee
Note the sign difference with respect to de Sitter; they are related by the analytic continuation  $\ell_{\rm AdS}\to i \ell_{\rm dS}$.  In Poincare coordinates the element $g$ is 
\be\label{poincare2}
g=\left[
\begin{array}{cc}
 \rho+ w\bar w/\rho & w/\rho  \\
\bar w/\rho& 1/\rho
\end{array}\right]~,
\ee
and the metric is
\bea
ds^2&=&{\ell^2\over2}{\rm Tr}(g^{-1}dg\,g^{-1}dg)
= {\ell^2\over \rho^2}\left(d\rho^2+dwd\bar w\right)~.
\eea

The identification that defines the BTZ  solution is
\be\label{quot2}
g\sim \gamma g \,\gamma^\dagger~,
\ee
with
\be
\gamma=\left[
\begin{array}{cc}
e^{-i\pi\tau} &0\\
0&e^{i\pi\tau}
\end{array}\right]~,\quad \tau=\tau_1+i\tau_2~.
\ee
This is the same element used for de Sitter, i.e. $\tau_{\rm AdS}=\tau_{\rm dS}$. In the coordinates \eqref{poincare}, the action of the identification \eqref{quot} implies
\be
\rho\sim e^{-i\pi (\tau-\bar\tau)} \rho~,\quad w\sim e^{-2\pi i\tau}w~.
\ee
A more convenient set of coordinates is given by
\bea\label{change2}
\rho&=&\left(|\tau|^2\over r^2+\tau_1^2\right)^{1/2}\exp(\tau_2\phi+\tau_1 t)~,\cr
w&=&\left(r^2-\tau_2^2\over r^2+\tau_1^2\right)^{1/2}\exp(-i\tau(\phi+i t))~,
\eea
where 
\be
\hat w=\phi+i t \sim \hat w +2\pi \sim \hat w + 2\pi \tau~,
\ee
in terms of which the metric reads
\be
{ds^2\over \ell_{\rm AdS}^2}={r^2dr^2\over(r^2-\tau_2^2)(r^2+\tau_1^2) }+r^2\left(d\phi +{\tau_1\tau_2\over r^2}dt\right)^2+{(r^2-\tau_2^2)(r^2+\tau_1^2)\over r^2}dt^2~.
\ee
The Euclidean BTZ black hole geometry described above is related to the dS${}_3$ cosmology \eqref{aa:2} by the analytic continuation $\ell_{\rm AdS}\to i\ell_{\rm dS}$ and $r\to ir$.


\section{Euclidean AdS$_3$ path integral and its analytic continuation}
In this appendix we collect the results for the Euclidean AdS$_3$ path integral computed in \cite{Maloney:2007ud}. The sum is given by \eqref{cc:3} which is divergent due to an infinite set of saddles with arbitrarily small volume in Planck units.  This divergence can be regulated in an $SL(2,\Z)$ invariant manner, giving 
\bea\label{regZAdS}
Z(\tau,\bar\tau)={1\over |\eta(\tau)|^2}&\Bigg[&E(2k-{1\over 12},0)+E(2k+2-{1\over 12},0)\cr &&-E(2k+1-{1\over 12},1)-E(2k+1-{1\over 12},-1)\Bigg]~
\eea
where $c_{\rm AdS}=24 k$. We have defined
\bea\label{es1}
 E(\kappa,\mu) =e^{2\pi (\kappa \tau_2+i \mu \tau_1)} +\sum_{n}e^{-2\pi i n \tau_1} E_n(\kappa,\mu,\tau_2)~,
 \eea
 with $\tau=\tau_1+i\tau_2$. The coefficients $E_n$ for $n\neq 0$ are given by
 \be\label{Enterms}
E_n(\kappa,\mu,\tau_2)=\sum_{m=0}^\infty I_{m,n}(\kappa,\mu,\tau_2)\chi_{m,n}(\mu) \tau_2^{-m}~.
\ee
where
\be
 I_{m,n}(\kappa,\mu,\tau_2)={(2\pi)^m\over m!}\int_{-\infty}^{\infty}dz\, e^{2\pi i n z \tau_2} (1+z^2)^{-m-1/2}(\kappa-i\mu z)^{m}~
\ee
and%
\be
\chi_{m,n}(\mu)=\sum_{c=1}^{\infty}c^{-2(m+1/2)}S(-n,\mu;c)~,\quad S(a,b;c)=\sum_{d\in (\Z/c\Z)^*}\exp\left({2\pi i\over c}(a d+ b d^{-1})\right)~.
\ee
For $n=0$ in \eqref{es1} we have
\bea\label{E00terms}
E_0(\kappa,\mu,\tau_2)=\sum_{m=0}^\infty w_m(\kappa,\mu) \tau_2^{-m}~,
\eea
where
\bea
w_m(\kappa,0)&=&{2^{m}\pi^{m+1/2}\zeta(2m)\over m\Gamma(m+1/2)\zeta(2m+1)}\kappa^m~,\cr
w_m(\kappa,\pm 1)&=&{2\pi^{m+1/2}\over m\Gamma(m+1/2)\zeta(2m+1)}T_m(\kappa)~,
\eea
and $T_m(x)$ is the Chebyshev polynomial of the first kind.

\begin{figure}
\centering
\includegraphics[width=0.7\textwidth]{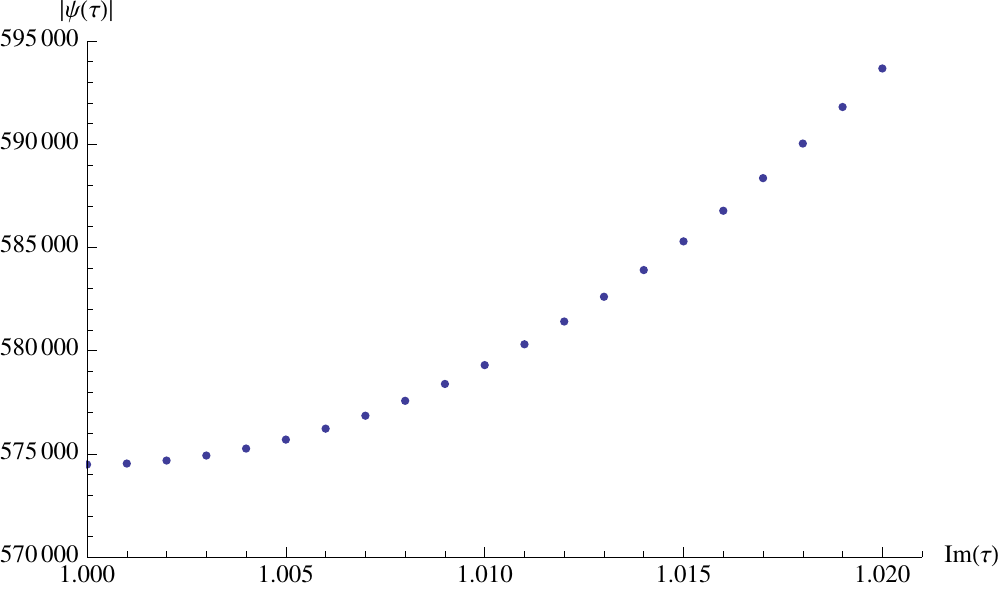}
\caption{The norm $|\Psi(\tau)|$ of the wave function \eqref{EPSI} for purely imaginary values of $\tau$ ($\tau_1=0$). In order to obtain stable numerical results it was necessary to include the ranges $n=-3\ldots 3$ and $m=0\ldots30$ in equations \eqref{E00terms}, \eqref{Enterms}. }
\label{numerics}
\end{figure}

The coefficients defining the regulated sum \eqref{regZAdS} are well defined for any complex $k$, so the regularization technique used in \cite{Maloney:2007ud} is sufficient to define the partition function in de Sitter space by analytic continuation in $k$. Taking $c_{\rm AdS}\to ic_{\rm dS}$, the wave function is
\bea\label{EPSI}
\Psi(\tau,\bar\tau)={1\over |\eta(\tau)|^2}&\Bigg[&E(2ik-{1\over 12},0)+E(2ik+2-{1\over 12},0)\cr &&-E(2ik+1-{1\over 12},1)-E(2ik+1-{1\over 12},-1)\Bigg]~.
\eea
To illustrate the shape of the wave function, in figure \ref{numerics} we plot the sum of first few hundred terms in the series expansion of \eqref{EPSI} for $c_{\rm dS}=2$ ($k=1/12$). Even though the coupling is small, the divergent behavior of $\Psi(\tau)$ is already evident from explicit evaluation of the first few terms in the sum. 


\begin{thebibliography}{99}

\bibitem{Hartle:1983ai}
  J.~B.~Hartle and S.~W.~Hawking,
  ``Wave Function of the Universe,''
  Phys.\ Rev.\ D {\bf 28} (1983) 2960.

\bibitem{Strominger:2001pn} 
  A.~Strominger,
  ``The dS / CFT correspondence,''
  JHEP {\bf 0110}, 034 (2001)
  [hep-th/0106113].
  
\bibitem{Maldacena:2002vr} 
  J.~M.~Maldacena,
  ``Non-Gaussian features of primordial fluctuations in single field inflationary models,''
  JHEP {\bf 0305}, 013 (2003)
  [astro-ph/0210603].
  
\bibitem{Witten:2001kn} 
  E.~Witten,
  ``Quantum gravity in de Sitter space,''
  hep-th/0106109.
 
\bibitem{Anninos:2012qw} 
  D.~Anninos,
  ``De Sitter Musings,''
  Int.\ J.\ Mod.\ Phys.\ A {\bf 27}, 1230013 (2012)
  [arXiv:1205.3855 [hep-th]].
  
\bibitem{Alishahiha:2004md} 
  M.~Alishahiha, A.~Karch, E.~Silverstein and D.~Tong,
  ``The dS/dS correspondence,''
  AIP Conf.\ Proc.\  {\bf 743}, 393 (2005)
  [hep-th/0407125].


\bibitem{McFadden:2009fg}
  P.~McFadden and K.~Skenderis,
  ``Holography for Cosmology,''
  Phys.\ Rev.\ D {\bf 81} (2010) 021301
  [arXiv:0907.5542 [hep-th]].
  
\bibitem{Sekino:2009kv} 
  Y.~Sekino and L.~Susskind,
  ``Census Taking in the Hat: FRW/CFT Duality,''
  Phys.\ Rev.\ D {\bf 80}, 083531 (2009)
  [arXiv:0908.3844 [hep-th]].
  
   
\bibitem{Maldacena:2011mk}
  J.~Maldacena,
  ``Einstein Gravity from Conformal Gravity,''
  arXiv:1105.5632 [hep-th].
  
\bibitem{Skenderis:2002wp} 
  K.~Skenderis,
  ``Lecture notes on holographic renormalization,''
  Class.\ Quant.\ Grav.\  {\bf 19}, 5849 (2002)
  [hep-th/0209067].

\bibitem{Gibbons:1976ue} 
  G.~W.~Gibbons and S.~W.~Hawking,
  ``Action Integrals and Partition Functions in Quantum Gravity,''
  Phys.\ Rev.\ D {\bf 15}, 2752 (1977).
  
\bibitem{Harlow:2011ke}
  D.~Harlow and D.~Stanford,
  ``Operator Dictionaries and Wave Functions in AdS/CFT and dS/CFT,''
  arXiv:1104.2621 [hep-th].
  

  
\bibitem{Witten:2010cx} 
  E.~Witten,
  ``Analytic Continuation Of Chern-Simons Theory,''
  arXiv:1001.2933 [hep-th].
  
\bibitem{Harlow:2011ny} 
  D.~Harlow, J.~Maltz and E.~Witten,
  ``Analytic Continuation of Liouville Theory,''
  JHEP {\bf 1112}, 071 (2011)
  [arXiv:1108.4417 [hep-th]].
  
\bibitem{Castro:2011xb} 
  A.~Castro, N.~Lashkari and A.~Maloney,
  ``A de Sitter Farey Tail,''
  Phys.\ Rev.\ D {\bf 83}, 124027 (2011)
  [arXiv:1103.4620 [hep-th]].
  
\bibitem{Castro:2011ke} 
  A.~Castro, N.~Lashkari and A.~Maloney,
  ``Quantum Topologically Massive Gravity in de Sitter Space,''
  JHEP {\bf 1108}, 040 (2011)
  [arXiv:1105.4733 [hep-th]].
  
\bibitem{Maloney:2007ud}
A.~Maloney, E.~Witten,
``Quantum Gravity Partition Functions in Three Dimensions,''
JHEP {\bf 1002}, 029 (2010).
{\tt [arXiv:0712.0155 [hep-th]]}.

\bibitem{Brown:1986nw} 
  J.~D.~Brown and M.~Henneaux,
  ``Central Charges in the Canonical Realization of Asymptotic Symmetries: An Example from Three-Dimensional Gravity,''
  Commun.\ Math.\ Phys.\  {\bf 104}, 207 (1986).
  
\bibitem{Maldacena:1998bw} 
  J.~M.~Maldacena and A.~Strominger,
  ``AdS(3) black holes and a stringy exclusion principle,''
  JHEP {\bf 9812}, 005 (1998)
  [hep-th/9804085].
  
 \bibitem{Dijkgraaf:2000fq}
R.~Dijkgraaf, J.M.~Maldacena, G.W.~Moore, E.P.~Verlinde,
``A Black hole Farey tail,''
{\tt [hep-th/0005003]}.
  
\bibitem{Witten:2007kt} 
  E.~Witten,
  ``Three-Dimensional Gravity Revisited,''
  arXiv:0706.3359 [hep-th].
  
\bibitem{Achucarro:1987vz} 
  A.~Achucarro and P.~K.~Townsend,
  ``A Chern-Simons Action for Three-Dimensional anti-De Sitter Supergravity Theories,''
  Phys.\ Lett.\ B {\bf 180}, 89 (1986).

\bibitem{Witten:1988hc} 
  E.~Witten,
  ``(2+1)-Dimensional Gravity as an Exactly Soluble System,''
  Nucl.\ Phys.\ B {\bf 311}, 46 (1988).
  
\bibitem{Vasiliev:1990en} 
  M.~A.~Vasiliev,
  ``Consistent equation for interacting gauge fields of all spins in (3+1)-dimensions,''
  Phys.\ Lett.\ B {\bf 243}, 378 (1990).


 \bibitem{Anninos:2011ui}
   D.~Anninos, T.~Hartman and A.~Strominger,
  ``Higher Spin Realization of the dS/CFT Correspondence,''
   arXiv:1108.5735 [hep-th].


\bibitem{Ouyang:2011fs} 
  P.~Ouyang, 
  ``Toward Higher Spin dS3/CFT2,''
  arXiv:1111.0276 [hep-th].

 \bibitem{Ng:2012xp}
   G.~S.~Ng and A.~Strominger,
   ``State/Operator Correspondence in Higher-Spin dS/CFT,''
  arXiv:1204.1057 [hep-th].

\bibitem{Das:2012dt} 
  D.~Das, S.~R.~Das, A.~Jevicki and Q.~Ye,
  ``Bi-local Construction of Sp(2N)/dS Higher Spin Correspondence,''
  arXiv:1205.5776 [hep-th].


\bibitem{Anninos:2012ft} 
  D.~Anninos, F.~Denef and D.~Harlow,
  ``The Wave Function of Vasiliev's Universe - A Few Slices Thereof,''
  arXiv:1207.5517 [hep-th].
  
\bibitem{Hertog:2011ky} 
  T.~Hertog and J.~Hartle,
  ``Holographic No-Boundary Measure,''
  JHEP {\bf 1205}, 095 (2012)
  [arXiv:1111.6090 [hep-th]].
  
\bibitem{Hartle:2012qb} 
  J.~B.~Hartle, S.~W.~Hawking and T.~Hertog,
  ``Accelerated Expansion from Negative $\Lambda$,''
  arXiv:1205.3807 [hep-th].
  
  
\bibitem{Balasubramanian:2001nb} 
  V.~Balasubramanian, J.~de Boer and D.~Minic,
  ``Mass, entropy and holography in asymptotically de Sitter spaces,''
  Phys.\ Rev.\ D {\bf 65}, 123508 (2002)
  [hep-th/0110108].

  
\bibitem{Klemm:2001ea}
  D.~Klemm,
  ``Some aspects of the de Sitter / CFT correspondence,''
  Nucl.\ Phys.\ B {\bf 625} (2002) 295
  [hep-th/0106247].
  
\bibitem{Bousso:2001mw}
  R.~Bousso, A.~Maloney and A.~Strominger,
  ``Conformal vacua and entropy in de Sitter space,''
  Phys.\ Rev.\ D {\bf 65} (2002) 104039
  [hep-th/0112218].

\bibitem{Giombi:2008vd} 
  S.~Giombi, A.~Maloney and X.~Yin,
  ``One-loop Partition Functions of 3D Gravity,''
  JHEP {\bf 0808}, 007 (2008)
  [arXiv:0804.1773 [hep-th]].
  
\bibitem{Kachru:2003aw} 
  S.~Kachru, R.~Kallosh, A.~D.~Linde and S.~P.~Trivedi,
  ``De Sitter vacua in string theory,''
  Phys.\ Rev.\ D {\bf 68}, 046005 (2003)
  [hep-th/0301240].
  
\bibitem{Banks:1984np} 
  T.~Banks, W.~Fischler and L.~Susskind,
  ``Quantum Cosmology In (2+1)-dimensions And (3+1)-dimensions,''
  Nucl.\ Phys.\ B {\bf 262}, 159 (1985).
  
\bibitem{Polyakov:2009nq} 
  A.~M.~Polyakov,
  ``Decay of Vacuum Energy,''
  Nucl.\ Phys.\ B {\bf 834}, 316 (2010)
  [arXiv:0912.5503 [hep-th]].
  
\bibitem{Polyakov:2012uc} 
  A.~M.~Polyakov,
  ``Infrared instability of the de Sitter space,''
  arXiv:1209.4135 [hep-th].
  
\bibitem{Krasnov:2000zq} 
  K.~Krasnov,
  ``Holography and Riemann surfaces,''
  Adv.\ Theor.\ Math.\ Phys.\  {\bf 4}, 929 (2000)
  [hep-th/0005106].
  
\bibitem{Yin:2007gv} 
  X.~Yin,
  ``Partition Functions of Three-Dimensional Pure Gravity,''
  Commun.\ Num.\ Theor.\ Phys.\  {\bf 2}, 285 (2008)
  [arXiv:0710.2129 [hep-th]].


\end{thebibliography}
\end{document}